\documentclass[a4paper]{article}
\usepackage{amsfonts}
\usepackage{amssymb}
\usepackage[margin=2cm]{geometry}
\usepackage[dvips]{graphicx}

\usepackage[T1]{fontenc}
\usepackage[latin2]{inputenc}

\usepackage{amsmath}
\usepackage{bbm}

\begin{document}


\title{Comment on "Resilient Quantum Computation in Correlated Environments: A Quantum Phase Transition Perspective" and "Fault-tolerant Quantum Computation with Longe-range Correlated Noise" }

\author{Robert Alicki \\ 
  {\small
Institute of Theoretical Physics and Astrophysics, University
of Gda\'nsk,  Wita Stwosza 57, PL 80-952 Gda\'nsk, Poland}\\
}

\date{\today}
\maketitle

In the two recent papers \cite{nov} , \cite{aha} following the earlier contributions \cite{novbar}
and \cite{ter}, respectively, the authors claim to provide general arguments supporting the idea of fault-tolerant quantum computations based on quantum error correction (QEC) schemes. However, the employed models of environment and the assumptions characterizing the qubit-environment coupling allow to decouple the quantum computer from the environment without any use of QEC. Remind that the damping rate of a periodic motion of the system with the frequency $\omega$ is given in the Born approximation by the spectral density of the reservoir 
\begin{equation}
R(\omega)= \int_{-\infty}^{\infty} e^{-i\omega t}\langle F(t)F\rangle_{bath}dt\geq 0
\label{spe}
\end{equation}
corresponding to the standard choice of the qubit-bath interaction $\sigma\otimes F$.
As the main frequency characterizing the computer is given by the inverse of the time needed to execute a quantum gate , i.e. $\omega_0 \simeq 1/t_0$, the average error per gate
scales as
\begin{equation}
\epsilon \simeq t_0 R(1/t_0)\ .
\label{err}
\end{equation}
The model proposed in \cite{nov} is essentially a linear coupling to a massless bosonic field, in $D$-spatial dimension, leading to the spectral density $R(\omega) \sim \omega^D $(at vacuum) or $\sim\omega^{D-1}$(at finite temperatures)  with the ultraviolet cut-off: $R(\omega) = 0$, for $|\omega|>\Lambda$. We can choose now one of the two strategies: "slow gates" $(t_0\to \infty; {\rm for}\  D>1\ {\rm or}\  D>2)$ or "fast gates"  $(t_0<< \Lambda^{-1})$ to eliminate the coupling to environment \cite{ali}. Similarly, the basic assumption of the papers \cite{aha},\cite{ter} means that
\begin{equation}
\frac{1}{2\pi}\int_{-\infty}^{\infty} R(\omega)\,d\omega =\langle F^2\rangle_{bath}\leq \|F\|^2\leq \frac{{\epsilon_0}^2}{{t_0}^2}
\label{ham}
\end{equation}
where $\epsilon_0 <<1$ is a certain fixed parameter. Here the condition (\ref{ham}) implies that $\lim_{|\omega|\to\infty}R(\omega) = 0$ and hence we can use again the "fast gates" strategy. Therefore, the results presented in both papers \cite{nov},\cite{aha} can hardly support the idea of  the "threshold theorems" which should work for the nonscalable, fixed error per gate. Here, in all papers \cite{nov}-\cite{ter}  the "fast gates" assumption is always present in a more or less hidden form. 

Physically, assumptions used in both papers are not generic. Beside the damping given by $R(\omega)\sim \omega^D$ we have always other mechanisms leading to "pure decoherence" with the rate $R(0)> 0$ (e.g. scattering processes described by the quadratic coupling to quantum fields). This allows only for a trade-off yielding certain minimal error per gate \cite{ali1}. Playing with ultraviolet cut-offs is also unphysical. They give only the restrictions on the energy regime where a given model is valid. For energies (frequencies) higher than the cut-off the other models of dissipation should be invoked. Generically, one expects that the full spectral density $R(\omega)$ globally increases with $|\omega|$ as more and more channels of the energy exchange become open. 

The interesting feature of the discussed models is the idea of "phase transition" related to the size of the coupling constant. Most likely it reflects the transition from the weak interaction regime, where the system preserves its identity and the perturbative formula (\ref{err}) is valid, to the  strong coupling regime where the non-perturbative "dressed system" description should be used (see e.g. rigorous results for the spin-boson model \cite{spo}).  
\par
\textbf{ Acknowledgements} This work is supported by the Polish Ministry
of Science and Information Technology- grant PBZ-MIN-008/P03/2003 and EC grant SCALA.


\begin{thebibliography}{99}

\bibitem{nov} E. Novais, E.R. Mucciolo and H.U. Baranger, Phys.Rev.Lett. {\bf 98}, 040501 (2007).

\bibitem{aha} D. Aharonov, A. Kitaev and J. Preskill, Phys.Rev.Lett. {\bf 96}, 050504 (2006).

\bibitem{novbar} E. Novais, and H.U. Baranger, Phys.Rev.Lett. {\bf 97}, 040501 (2006).

\bibitem{ter} B.M. Terhal and G. Burkard, Phys.Rev.{\bf A 71}, 012336 (2005).

\bibitem {ali} R. Alicki , Chem. Phys. {\bf 322}, 75 (2006); Open Sys.\& Information Dyn. {\bf 11}, 53 ,(2004).

\bibitem{ali1} R. Alicki, M. Horodecki, P. Horodecki, R. Horodecki, L. Jacak, and P. Machnikowski, Phys. Rev.{\bf A 70}, 010501(R), (2004). 

\bibitem {spo} H. Spohn, Commun.Math.Phys.{\bf 123}, 277 (1989).

\end{thebibliography}
\end{document}